\documentclass{article}  
\usepackage{jamaica04}
\usepackage{epsfig}
\frompage{245} \topage{252}                                              
\def\rightmark{Can the QGP be found using only an EMCal?}
\def\leftmark{M.J.Tannenbaum for the PHENIX Collaboration}
 
\title{Can the QGP be found \\using only an Electromagnetic Calorimeter?} 
\authors{
{M. J. Tannenbaum $^1$ for the PHENIX Collaboration %
}\\[2.812mm]
{\normalsize
\hspace*{-8pt}$^1$ Brookhaven National Laboratory, \\ 
Upton, NY 11973-5000, USA\\
Supported by the U. S. Dept. of Energy under contract DE-AC02-98CH10886.\\[0.2ex] 
}}
 
\abstract{Measurements using the finely segmented EM Calorimeter in PHENIX are presented. The issue of whether these are sufficient to claim discovery of the Quark Gluon Plasma is discussed.}

 
\begin{document}
 
\maketitle
\setcounter{page}{1}

\section{How to discover the Quark Gluon Plasma}\label{sec:howto}

	The ``gold-plated" signature for the discovery of the Quark Gluon Plasma and the proof of ``deconfinement" is the suppression of $J/\Psi$ due to "Debye screening" which dissociates the bound pair of $c-\bar{c}$ quarks.~\cite{MatsuiSatz} The PHENIX experiment at RHIC is a special purpose detector~\cite{PHENIXdet} designed to observe this classic signature, and, more generally, to measure rare processes involving leptons and photons at the highest luminosities. 
\section{How the PHENIX detector was designed}
	
\begin{figure}[!t]
\begin{tabular}{cc}
\centering\psfig{file=fig1pi2grb.epsf,width=2.8in}
\hspace{0.014in}
\centering\psfig{file=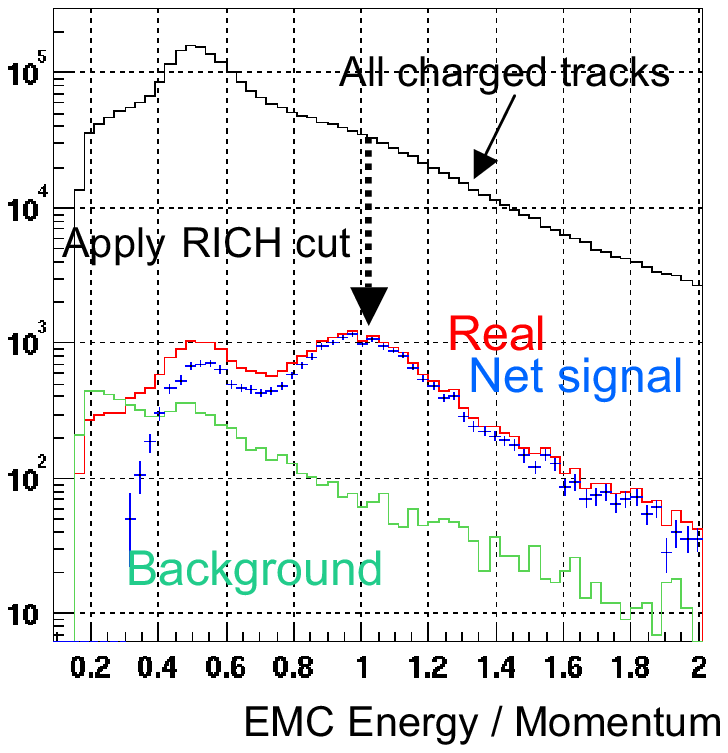,width=2.0in}
\end{tabular}\vspace*{-0.12in}
\caption[]{(left) Schematic of $\pi^{\pm}$, $e^{\pm}$ and $\gamma$ in PHENIX, with ElectroMagnetic Calorimter (EMCal) and Ring Imaging Cherenkov Counter (RICH). (right) Energy/momentum for all charged particles detected in the EMcal with and without a RICH signal.}
\label{fig:pxworks}\vspace*{-0.12in}
\end{figure}

	The principles of the design of PHENIX are illustrated in Fig.~\ref{fig:pxworks}. 
In order to detect electrons in hadron collisions, where the typical ratio of $e^{\pm}/\pi^{\pm}$ is known to be of order of $10^{-4}$ for prompt electrons (from charm) at $23.5 \leq \sqrt{s}\leq 62.4$ GeV~\cite{CCRS} (ISR energies), one must plan on a charged pion rejection of $ > 10^5$. PHENIX decided to use a Ring Imaging Cerenkov counter (RICH) in combination with an Electromagnetic Calorimeter (EMCal) to achieve this rejection. There is also the issue of the huge background of $e^{\pm}$ from internal and external conversion of the photons from $\pi^0$ decay or from direct photon production, which must be measured and understood to high precision. The EMCal is crucial for this purpose.  
	The EMCal measures the energy of $\gamma$ and $e^{\pm}$ and reconstructs $\pi^0$ from 2 photons. It measures a decent time of flight (TOF), 0.3 nanoseconds over 5 meters, allowing photon and charged particle identification. A high precision TOF over part of the aperture allows improved charged hadron identification. Electrons are identified by a count in the RICH and matching Energy and momentum ($E/p$), where the momentum is measured by track chambers in a magnetic field. Charged hadrons deposit only minimum ionization in the EMCal ($\sim 0.3$ GeV), or higher if they interact, and don't count in the RICH ($\pi^{\pm}$ threshold 4.7 GeV/c). Thus, requiring a RICH signal rejects all charged hadrons with $p < 4.7$ GeV/c, leaving only $e^{\pm}$ as indicated by the $E/p=1$ peak in Fig.~\ref{fig:pxworks}-right. $\pi^{\pm}$ can be identified  above 4.7 GeV/c by a RICH signal together with an EMCal hit with minimum ionization or greater. 
	It is amusing to realize that once you decide to measure electrons, you must make an excellent $\pi^0$ measurement to understand the background, and this implies a detector which can measure and identify almost all particles, as exemplified by the PHENIX central spectrometer.

	Admittedly, in order to detect the QGP, there must be charged particle tracking to complement the EMCal and measure the momenta of charged hadrons and $e^{\pm}$. In the following,  we discuss how an EMCal can be used to measure centrality, fluctuations, high $p_T$ pions and direct photon production. Further discussion on $J/\Psi$ and open charm detection in PHENIX using single $e^{\pm}$ and $e^+ e^-$ pairs is given in a related  article.~\cite{MJTQCD2003}
	\section{Centrality Measurement and Fluctuations}
	\begin{figure}[hbt]
\centering\psfig{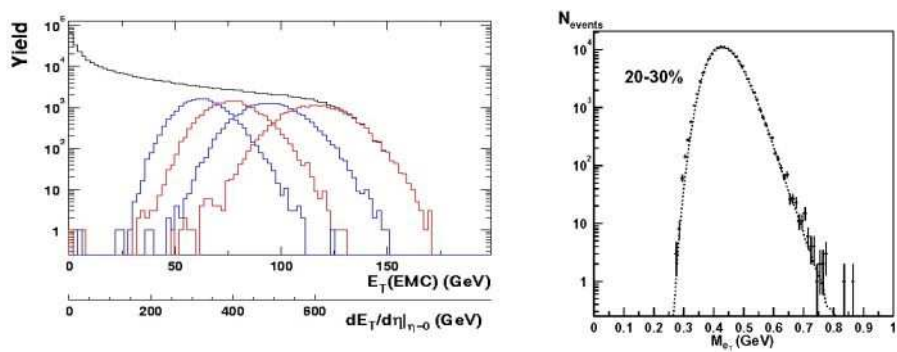}
\vspace*{-0.12in}
\caption[]{(left) PHENIX $E_T$ distribution in Au+Au collisions at $\sqrt{s_{NN}}=200$ GeV showing centrality selections of 0-5\%, 5-10\%, 10-15\%, 15-20\%. (right) Event-by-event average $E_T$/cluster in the EMCal for 20-30\%  centrality (data points), random baseline (dotted line).}
\label{fig:ETfluc}
\end{figure}
    $E_T$ distributions at mid-rapidity (see Fig.~\ref{fig:ETfluc}-left), measured in either hadron or EM calorimeters, have traditionally been used as a measure of centrality in B+A collisions in terms of the upper percentiles of the distribution.~\cite{MJTETreviews} They also measure the Bjorken estimator of the energy density of the collision:~\cite{Bj}
    \begin {equation}
    \epsilon_{Bj}={1\over {\tau_0 \pi R^2}} {{d E_T} \over dy} \qquad , 
    \label{eq:eBj}
     \end{equation}
which is 5.5 GeV/fm$^3$ for Fig.~\ref{fig:ETfluc}-left, well above the value of 1 GeV/fm$^3$ nominally required for QGP formation. 
Fluctuations can also be studied. For instance, is the shape of the upper edge of the centrality selected $E_T$ distributions in Fig.~\ref{fig:ETfluc}-left random, or is it evidence of non-random fluctuations? The event-by-event distribution $M_{e_T}$ of the average $E_T$ per cluster, which is closely related to $E_T$ and also to $M_{p_T}$, the event-by-event average $p_T$ of charged particles, is shown in Fig.~\ref{fig:ETfluc}-right.~\cite{PXPRC66} This measures the non-random fluctuations by comparison to a random baseline from mixed events.  $M_{p_T}$ is defined: 
 \begin{equation}
M_{p_T}=\overline{p_T}={1\over n} \sum_{i=1}^n p_{T_i}={1\over 
n} E_{Tc} \qquad ,\label{eq:defMpT}
\end{equation}
where $E_{Tc}$ is the analog of $E_T$ for charged particles. 
PHENIX has shown that the non-random $M_{p_T}$ fluctuations are of the order of 1\%~\cite{PXppg027} and are consistent with being the result of momentum correlations from jets. This places severely small limits on the critical fluctuations expected for a first order phase transition. 

\section{Hard-Scattering as a new probe of the QGP} 
    One sure way to get a partonic probe into the dense nuclear matter produced in an A+A collision is to use the high $p_T$ partons produced by hard-scattering. For p-p collisions in the RHIC energy range, hard-scattering is the dominant process of particle production with $p_T \geq 2$ GeV/c at mid-rapidity. Particles with $p_T\geq 2$ GeV/c are
produced from states with two roughly back-to-back jets which are
the result of scattering of constituents of the nucleons as described
by QCD. 
    
\begin{figure}[!hbt]
\begin{tabular}{cc}
\centering\psfig{file=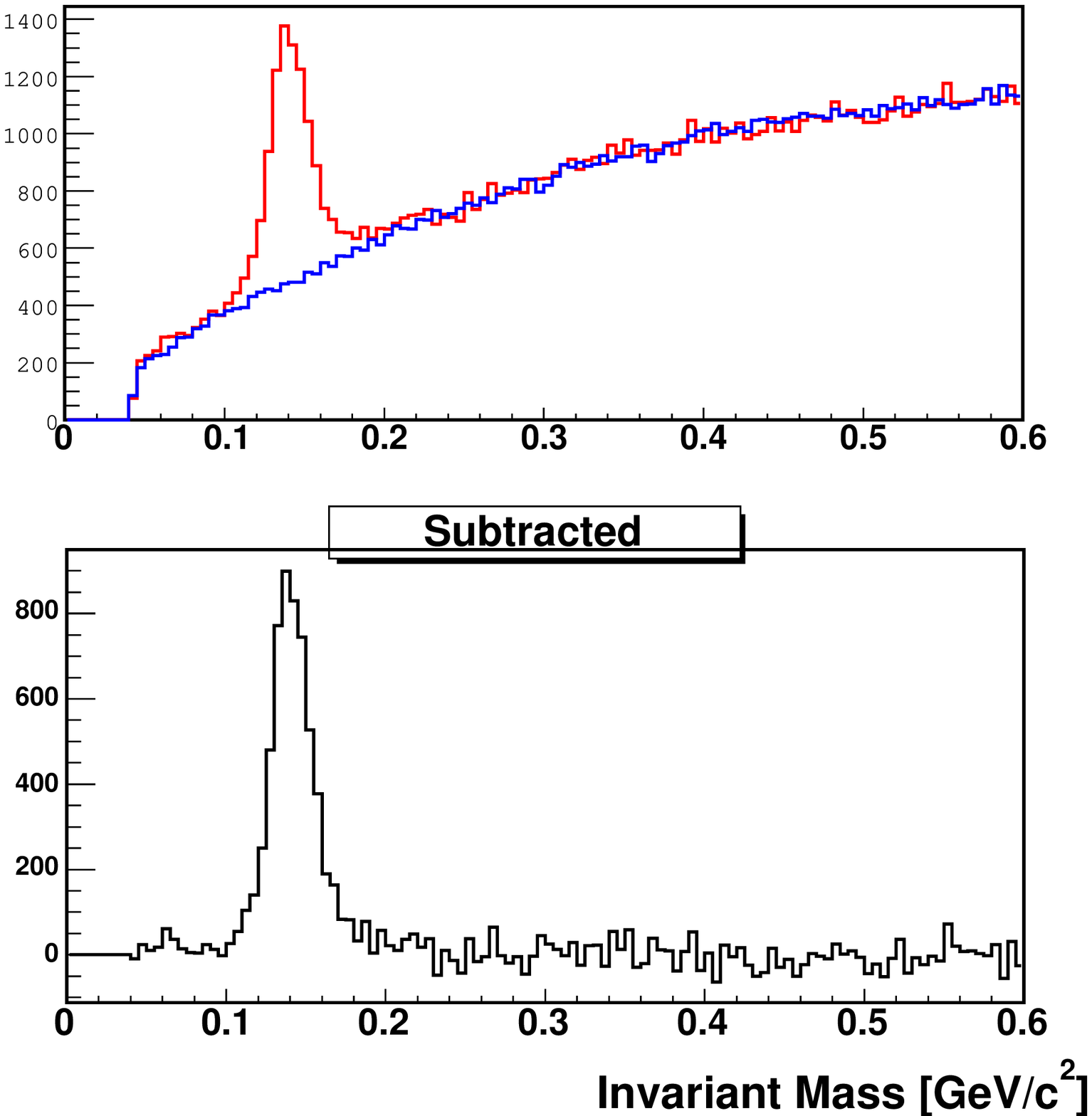,width=2.5in}
\hspace{0.014in}
\centering\psfig{file=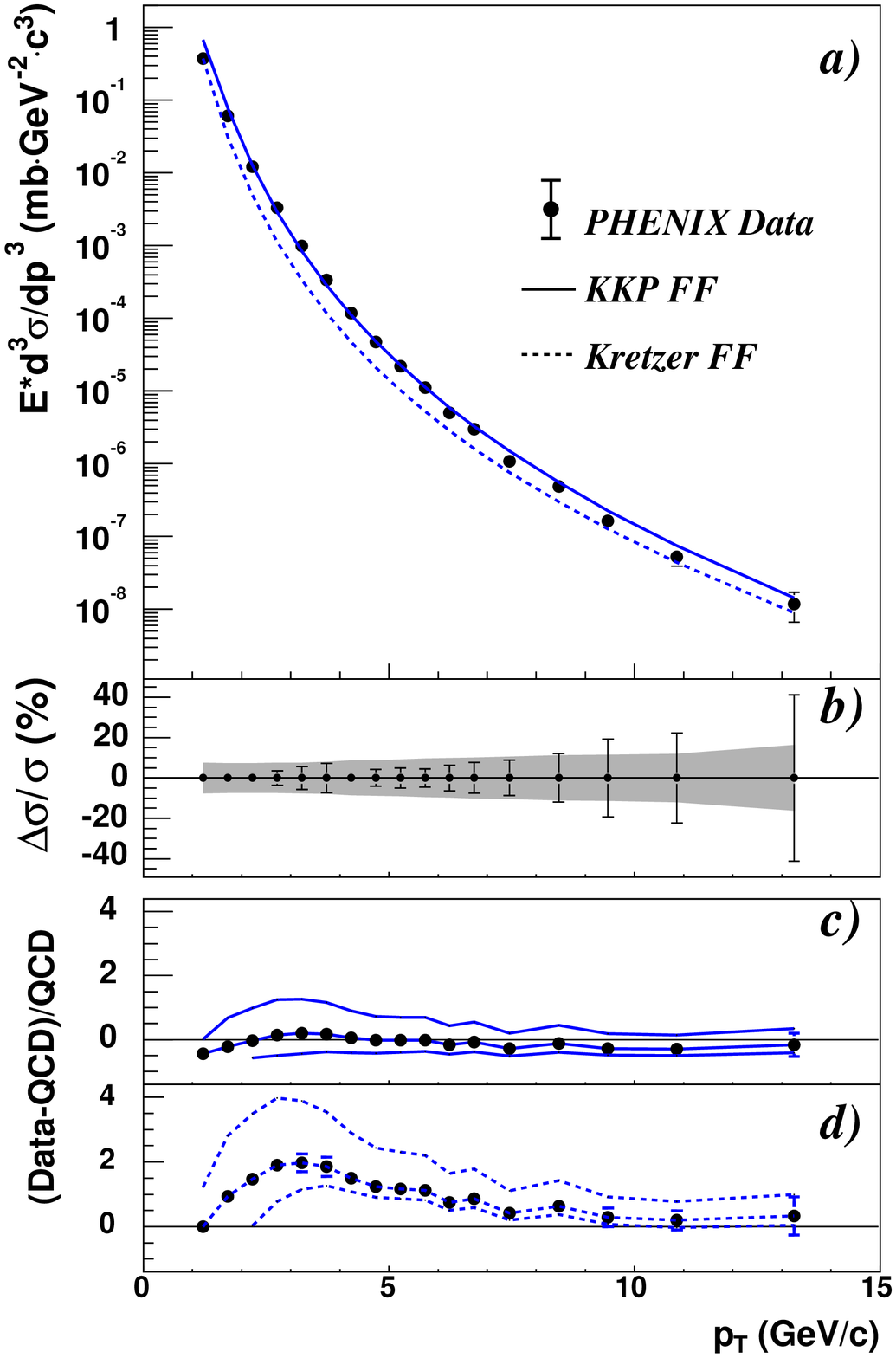,width=2.3in}
\end{tabular}\vspace*{-0.12in}
\caption[]{(left) Pizero peak Au+Au minimum bias collisions for $4.5\leq p_T\leq 5.0$ GeV/c. (right) PHENIX $\pi^0$ invariant cross section at mid-rapidity from p-p collisions at $\sqrt{s}=200$ GeV, together with NLO QCD predictions.}
\label{fig:pizpp}
\end{figure}

    The high granularity of the individual EMCal towers is $(\delta\eta \times \delta\phi)$ $\sim (0.01\times 0.01)$, which allows the two photons from a $\pi^0$ to be resolved from a single photon cluster for values of $p_T$  up to $30$ GeV/c, or higher.  The performance of the EMCal for detecting $\pi^0\rightarrow \gamma +\gamma$ in Au+Au collisions is shown in Fig.~\ref{fig:pizpp}-left. The $\pi^0$ cross section measurement from p-p collisions at $\sqrt{s}=200$ GeV (Fig.~\ref{fig:pizpp}-right) nicely agrees with NLO QCD predictions over the range $2.0\leq p_T \leq 15$ GeV/c.~\cite{PXpizpp200}   
\subsection{Scaling hard-scattering from p-p to p+A and B+A collisions} 
\begin{figure}[!hbt]
\begin{tabular}{cc}
\centering\psfig{file=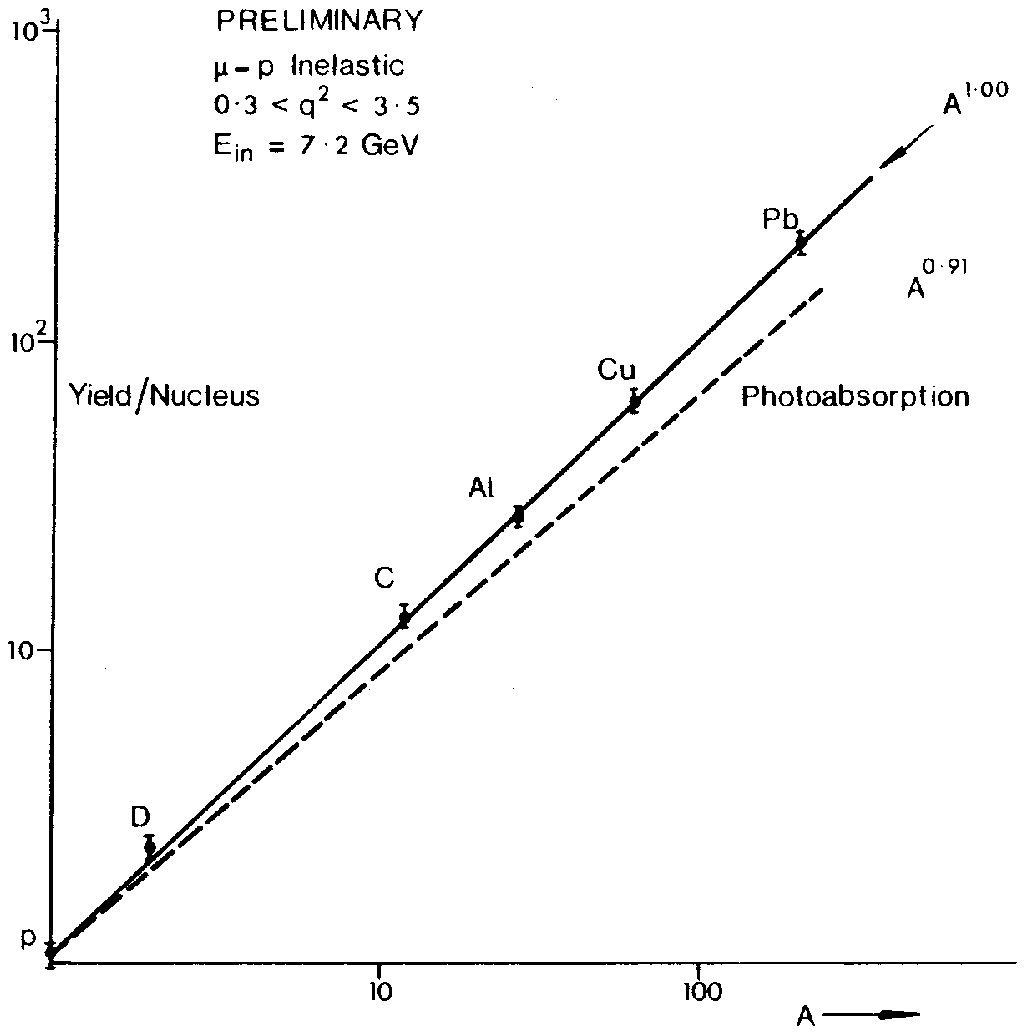,width=2.4in}
\vspace*{0.24in}
\hspace{0.014in}
\centering\psfig{file=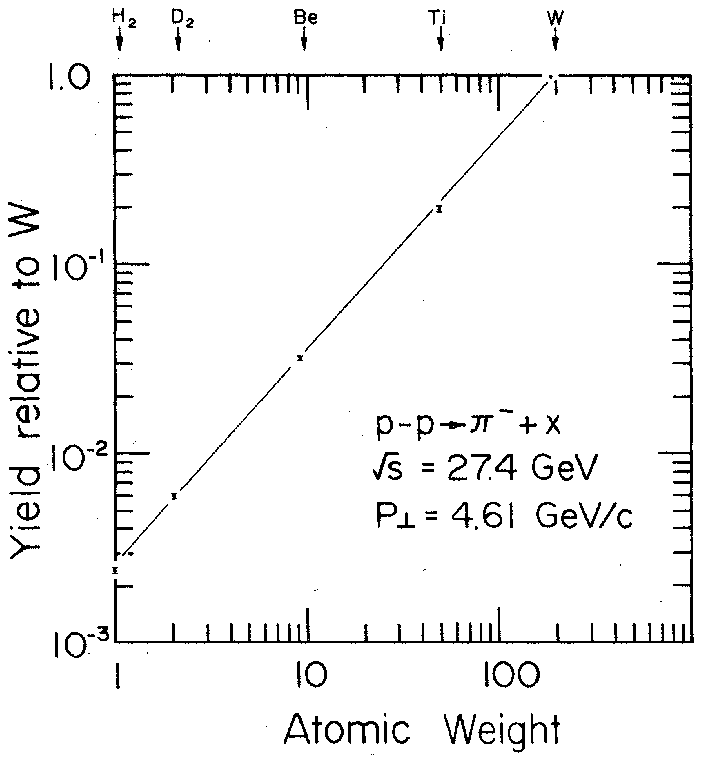,width=2.4in}
\end{tabular}\vspace*{-0.12in}
\caption[]{$A$ dependence of $\mu$-A scattering (left) compared to p+A (right).}
\label{fig:muApA}
\end{figure}

    Since hard-scattering is pointlike, with distance scale $1/p_T \leq 0.1$ fm, the cross section in p+A or B+A collisions, compared to p-p, is simply proportional to the relative number of possible pointlike encounters, or $A$ $(BA)$, for p+A (B+A) minimum bias collisions, and $T_{AB}(b)$, the overlap integral of the nuclear profile functions, as a function of impact parameter, $b$.  In lepton scattering, where hard-scattering was discovered,~\cite{SLAC-MIT} the cross section for $\mu$-A collisions is indeed proportional to $A^{1.00}$ (Fig~\ref{fig:muApA}-left).~\cite{MMay} In p+A collisions, the cross section at a given $p_T$ also scales as a power law, $A^{\alpha (p_T)}$ (Fig~\ref{fig:muApA}-right), but the power ${\alpha (p_T)}$ is greater than 1! This is the famous ``Cronin Effect".~\cite{Cronin}  
 
\section{Measurements of high $p_T$ $\pi^0$ and $\pi^{\pm}$ production in d+Au and Au+Au collisions}
\begin{figure}[!hbt]
\begin{tabular}{cc}
\centering\psfig{file=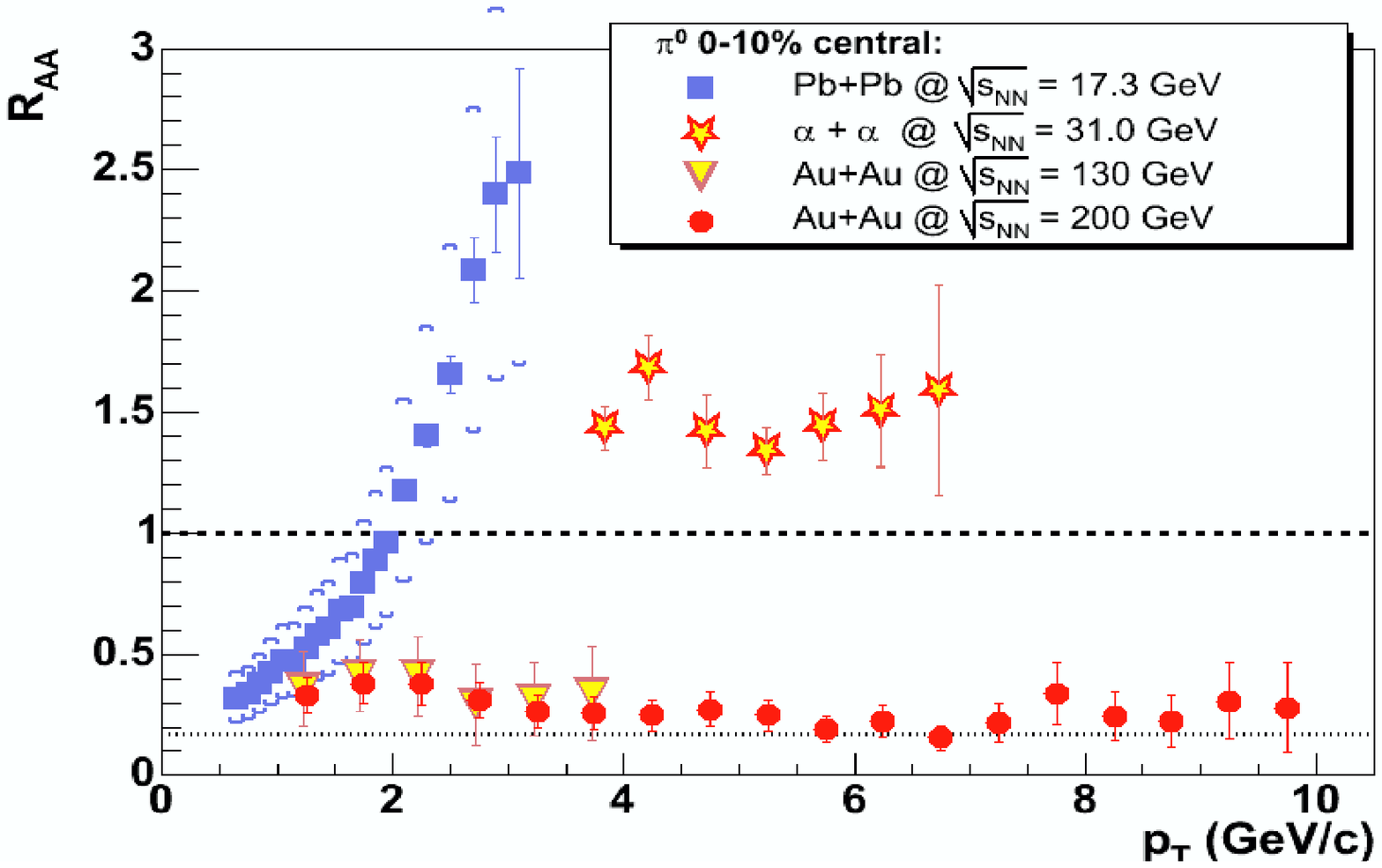,width=2.7in}
\hspace*{0.014in}
\centering\psfig{file=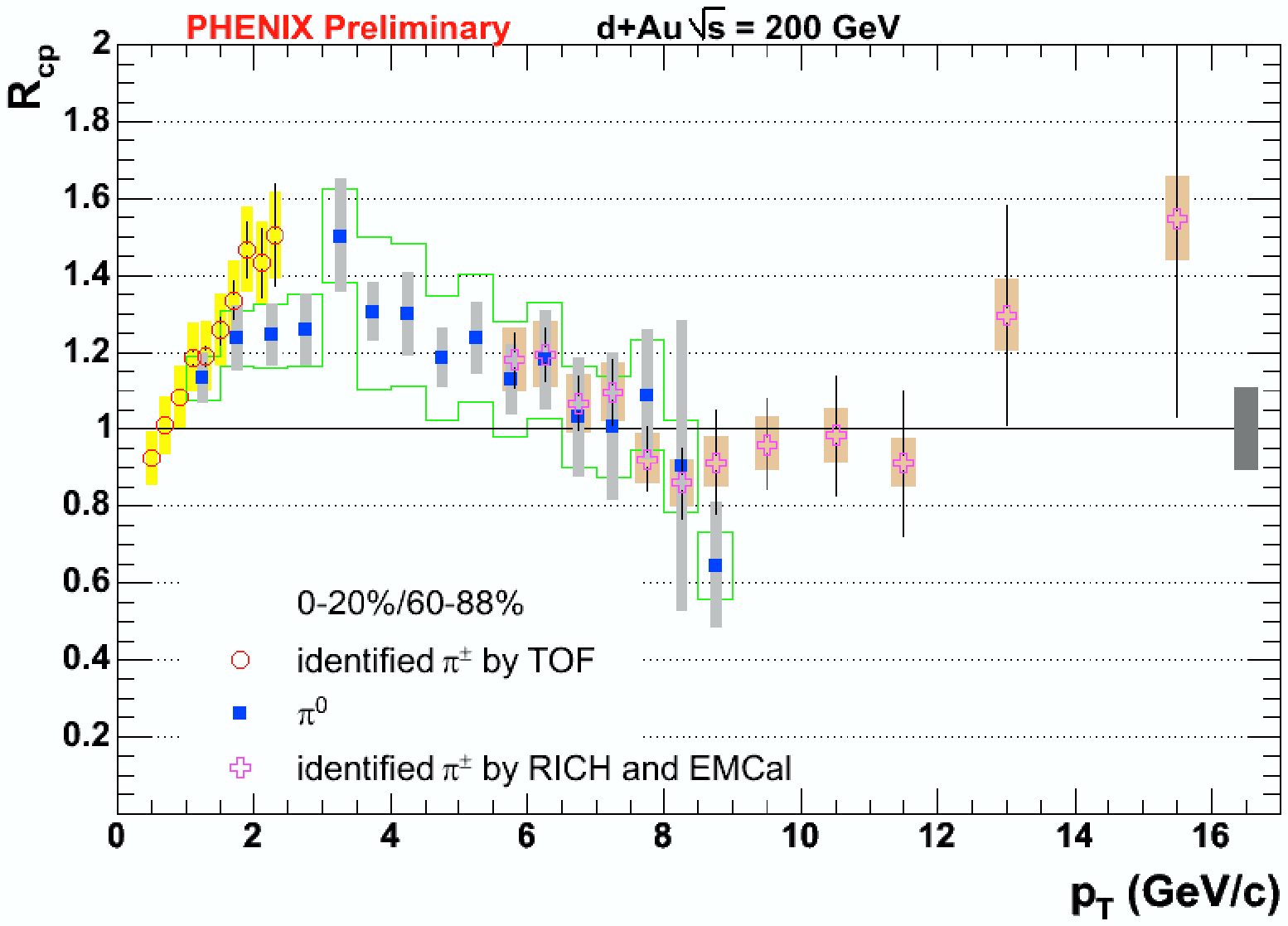,width=2.05in}
\end{tabular}\vspace*{-0.12in}
\caption[]{(left) The nuclear modification factor $R_{AA}$ for high $p_T$ $\pi^0$ production in A+A collisions. (right) $R_{CP}$ (point-like-scaled central/peripheral) for pions in d+Au collisions at RHIC showing a Cronin effect.~\cite{DdEQM2004}}
\label{fig:pizAA}
\end{figure}

    I believe that the most spectacular discovery at RHIC so far is the suppression of high $p_T$ $\pi^0$ in Au+Au collisions (Fig.~\ref{fig:pizAA}-left).~\cite{discovery,PXpizAuAu200,chdifferent}  

In Au+Au central collisions at $\sqrt{s_{NN}}=$130 GeV~\cite{discovery} and 200 GeV~\cite{PXpizAuAu200} the ratio, $R_{AA}$, of the Au+Au $\pi^0$ cross-section to the point-like-scaled p-p cross section is pretty much the same as a function of $p_T$ for both 130 and 200 GeV, and in each case is much less than 1! The high $p_T$ $\pi^0$ from  central Au+Au collisions are suppressed by a factor of from 3-5 in the range $2\leq 10$ GeV/c! This ``anti-Cronin" effect is truly astounding and unique at RHIC. $\pi^0$ production in A+A collisions at lower $\sqrt{s_{NN}}\leq 31$ GeV
and in p+A (or d+A) collisions, even at RHIC~\cite{PXdAu} (Fig.~\ref{fig:pizAA}-right) all show a Cronin effect, an enhancement with respect to point-like-scaled p-p collisions.  
This implies that the suppression in Au+Au collisions at RHIC energies is due to final state interaction with the hot, dense and possibly deconfined medium produced, and is not due to an initial state effect such as gluon shadowing or coherence which would also show up in d+Au collisions.  
\subsection{The power of the power-law.} 
\begin{figure}[!thb]
\begin{tabular}{cc}
\centering\psfig{file=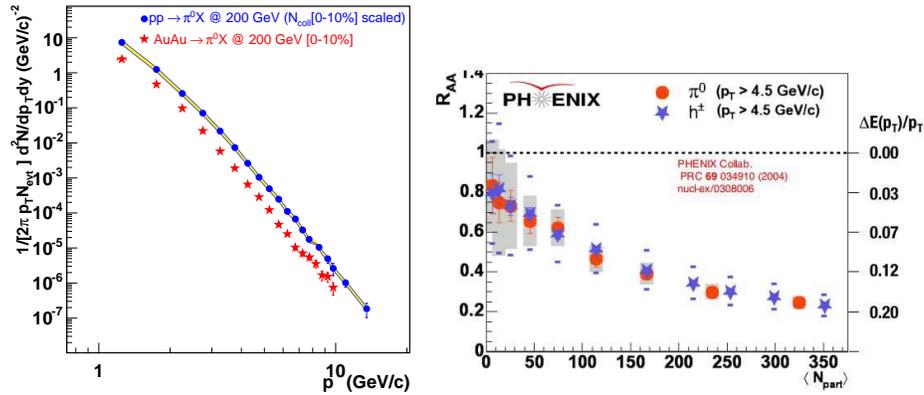,width=2.1in}
\hspace*{0.014in}
\centering\psfig{file=dEvsCentgrb.epsf,width=2.6in}
\end{tabular}\vspace*{-0.12in}
\caption[]{(left) log-log plot of PHENIX $\pi^0$ $p_T$ spectrum in central Au+Au collisions compared to scaled p-p data. (right) Centrality dependence of $R_{AA}$ for $p_T \geq 4.5$ GeV/c with computed $\Delta E(p_T)/p_T$.}
\label{fig:dEdAu}
\end{figure}

    A power-law dependence is characteristic of hard-scattering. This is shown for both the Au+Au central and p-p data in Fig.~\ref{fig:dEdAu}-left.
The invariant cross section behaves as a pure power,  $p_T^{-n}$,  for $p_T\geq 3$ GeV/c,  with $n=8.1\pm0.1$. It is easy to see that the medium effect in Au+Au collisions is equivalent to either a suppression of a factor $R_{AA}$ at a given $p_T$ or a shift in $p_T$ of the Au+Au spectrum relative to the p-p spectrum. In fact the spectra are quite parallel, so the shift to match the spectra is a constant for all $p_T$, corresponding to a constant $R_{AA}$, or to a constant fractional shift 
$\Delta E(p_T)/p_T=1-R_{AA}(p_T)^{1/(n-1)}$ 
for $p_T\geq 4.5$ GeV/c, for all centralities.~\cite{BDMS} (See Fig.~\ref{fig:dEdAu}-right.)   

	Another advantage of the power-law is that it makes the precise calculation of the photon spectrum from $\pi^0\rightarrow \gamma +\gamma$ decay rather easy. 
	\begin{equation}
\mbox{If}\qquad {{dn_{\pi^0}}\over{p_T dp_T}}\propto p_T^{-n}\qquad \mbox{then}\qquad 
\left . {\gamma\over \pi^{0}}\right |_{\pi^0}\!\! (p_T)=2/(n-1) \qquad. 
\label{eq:powerlaw}
\end{equation}
Since the the photons from $\pi^0\rightarrow \gamma+\gamma$, $\eta\rightarrow \gamma+\gamma$ and similar decays are the principal background to direct photon production, the importance of a precise estimate of this background can not be overstated. 

    	\section{Direct Photon production in Au+Au}
\begin{figure}[!hbt]
\begin{tabular}[b]{cc}
\begin{tabular}[b]{c}
\centering\psfig{file=bkg-gampigrb.epsf,width=2.0in}\cr
\centering\psfig{file=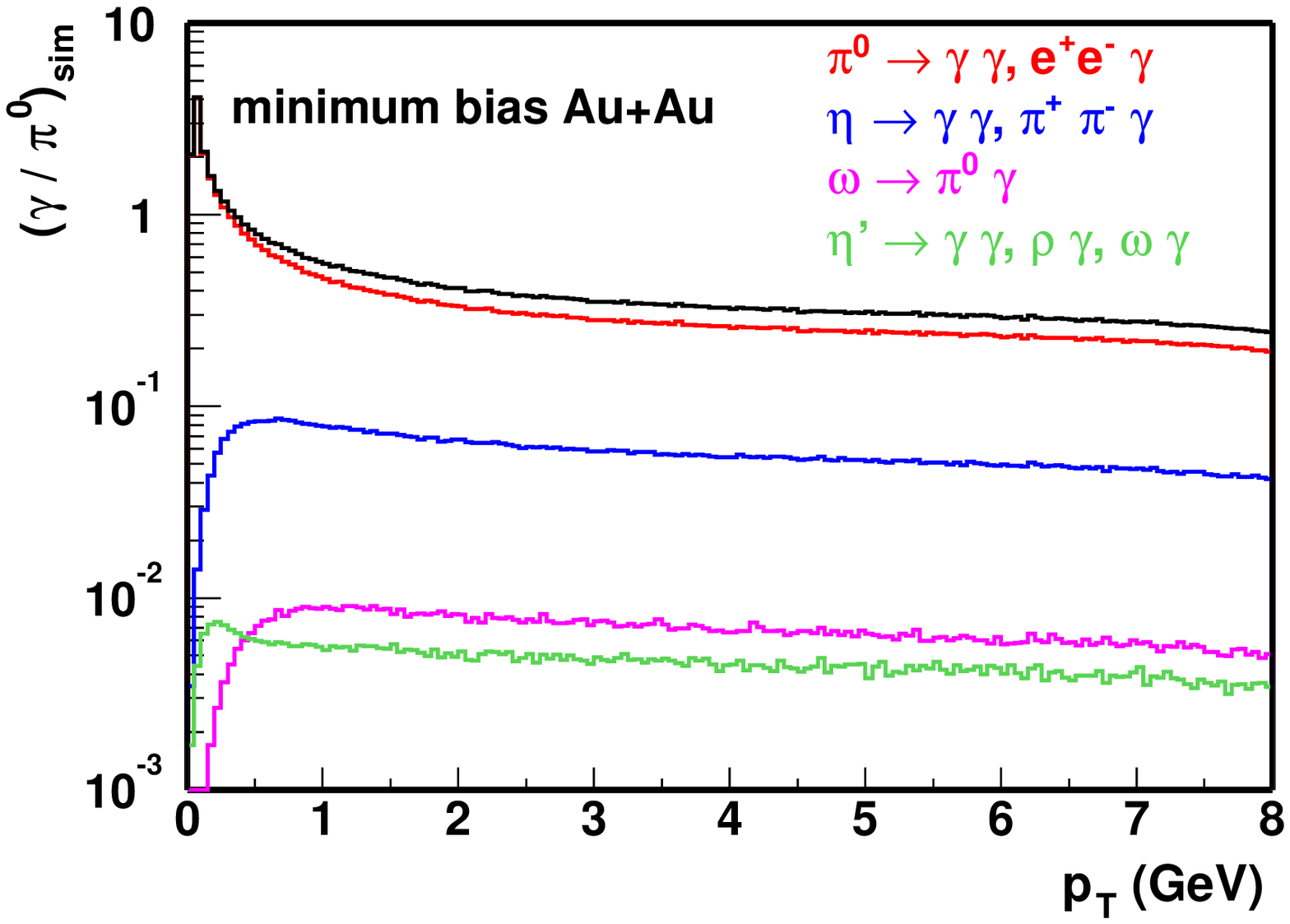,width=2.0in}
\end{tabular}
\hspace*{-0.12in}
\centering\psfig{file=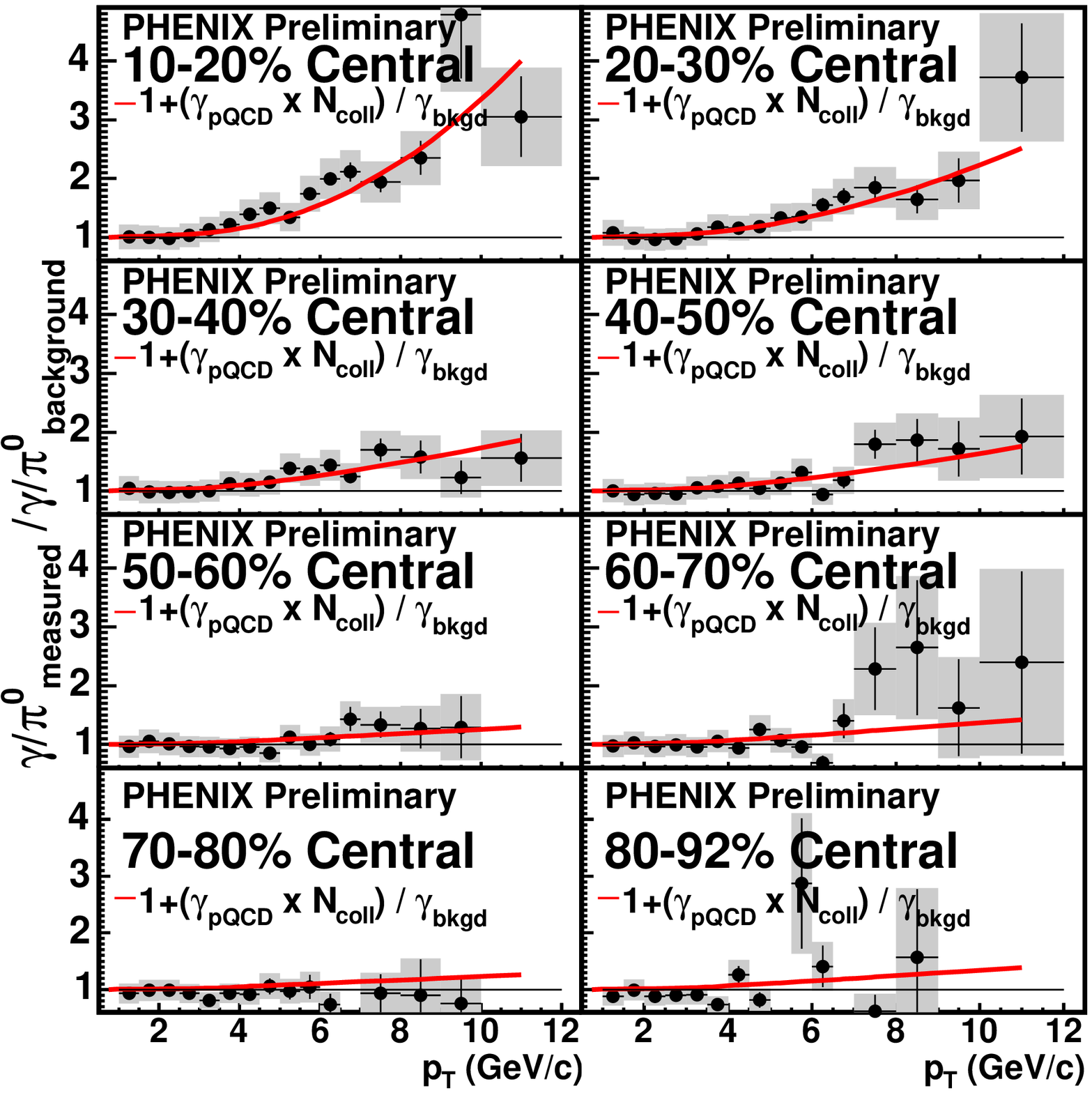,width=2.8in}
\end{tabular}\vspace*{-0.12in}
\caption[]{(left)-(top) Calculation of background photon spectrum from known decays, expressed as $\gamma|_{\rm background}/\pi^0$ compared to Eq.~\ref{eq:powerlaw}. (left)-(bottom) Individual components of the background. (right) PHENIX direct photon measurements relative to the calculated background for $\sqrt{s_{NN}}=200$ GeV Au+Au collisions, as a function of centrality (data points). The curves represent a pQCD calculation of direct photons in p-p collisions scaled to Au+Au assuming pure point-like ($N_{\rm coll}$) scaling, with no suppression. }
\label{fig:dirphoton}
\end{figure}

The first measurement of direct photon production in Au+Au collisions at RHIC was shown by the PHENIX collaboration at Quark Matter 2004.~\cite{JFrantz} These photons originate from the  `hard' process $g+q\rightarrow \gamma +q$. There are lots of photons at RHIC. The problem is to subtract the background photons from $\pi^0$ and $\eta$ decay. The simple (but accurate) approach in Eq.~\ref{eq:powerlaw} gives, for $n=8.1$, $\gamma|_{\pi^0}/\pi^0\sim 1.2\times 2/7.1=0.34$ where  the factor 1.2 includes $\eta\rightarrow \gamma+\gamma$ (estimated). The validity of this simple approach is evident by comparison to the full calculation of $\gamma/\pi^0$ from known decays (Fig.~\ref{fig:dirphoton}-left). 
 
 	The direct photon photon measurements in Au+Au as a function of centrality follow pure point-like ($N_{\rm coll}$) scaling relative to a pQCD calculation for p-p collisions, with no suppression, quite different from the $\pi^0$ measurements. This is good news because photons should not interact with the hot, dense medium. Ironically, the fact that the main source of background, the $\pi^0$'s, are suppressed in Au+Au collisions, while the direct photons are not, makes the measurement easier in Au+Au collisions than in p-p collisions.
	\section{Conclusions}
	It may very well be possible to find the QGP using only an EM Calorimter. But it's better with $e^{\pm}$ i.d. at the trigger level, and still better with $\mu^{\pm}$, too. It is even better with full charged hadron identification. The only remaining question, since $\pi^0$ suppression is spectacular but is apparently not sensitive to deconfinement~\cite{BDMS}, is whether we have found the QGP at RHIC, or merely the densest, nuclear matter ever to have existed since the Big Bang?  

\vfill\eject

\begin{thebibliography}{99}
\bibitem{chdifferent} Non-identified charged particles and identified protons have a quantitatively different behavior which is also very interesting. See the talk by J. Velkovska in these proceedings.
\end{thebibliography}
\end{document}